%% file: MAIN.tex
\documentclass[sigconf]{acmart}
\usepackage{soul}
\usepackage{booktabs}
\usepackage{array}
\usepackage{lscape}
\usepackage{tcolorbox}
\usepackage{cleveref}
\usepackage{caption}
\usepackage{tabularx}
\newcolumntype{b}{X}
\newcolumntype{s}{>{\hsize=.5\hsize}X}
\renewcommand{\hl}[1]{#1}

\AtBeginDocument{%
  }

\newtcolorbox{mybox}[1]{%
    tikznode boxed title,
    enhanced,
    arc=0mm,
    interior style={white},
    attach boxed title to top center= {yshift=-\tcboxedtitleheight/2},
    fonttitle=\bfseries,
    colbacktitle=white,coltitle=black,
    boxed title style={size=normal,colframe=white,boxrule=0pt},
    title={#1}}

\renewenvironment{quote}%
  {\list{}{\leftmargin=0in\rightmargin=0in}\item[]}%
  {\endlist}

\copyrightyear{2025}
\acmYear{2025}
\setcopyright{cc}
\setcctype{by}
\acmConference[CHI '25]{CHI Conference on Human Factors in Computing Systems}{April 26-May 1, 2025}{Yokohama, Japan}
\acmBooktitle{CHI Conference on Human Factors in Computing Systems (CHI '25), April 26-May 1, 2025, Yokohama, Japan}\acmDOI{10.1145/3706598.3713470}
\acmISBN{979-8-4007-1394-1/25/04}

\begin{document}

% \title[Objection, your Honour, the LLM Told Me to]{Objection, your Honour, the LLM Told Me to:\\An Examination of Lay-People's Willingness to Act on Legal Advice Generated by a Large Language Model (LLM)}

% \title{Objection Overruled! Lay People can Distinguish the Source of Legal Advice but Favour the Large Language Model (LLM)}

% \title{Objection Overruled! Lay People can Distinguish Large Language Models (LLM) from Lawyers, but still Favour Advice from LLMs}

% \title{Objection, your Honour, the LLM Told Me to:\\ Lay People can Distinguish Large Language Models from Lawyers, but still Favour Advice from an LLM}

\title{Objection Overruled!\\ Lay People can Distinguish Large Language Models from Lawyers, but still Favour Advice from an LLM}

%\title[Objection, your Honour, the LLM Told Me to]{Objection, your Honour, the LLM Told Me to:\\An Examination of Lay-People's Willingness to Act on, and Identify the Source of, Legal Advice Generated by a Large Language Model (LLM)}

%\title{I Rest my Case with the Aid of this LLM:\\An Examination of Lay-People's Willingness to Act on Legal Advice Generate by an Large Language Model (LLM)}

\author{Eike Schneiders}
\affiliation{%
  \institution{University of Southampton}
  \city{Southampton}
  \country{UK}}
\email{eike.schneiders@soton.ac.uk}

\author{Tina Seabrooke}
\affiliation{%
  \institution{University of Southampton}
  \city{Southampton}
  \country{UK}}
\email{t.seabrooke@soton.ac.uk}

\author{Joshua Krook}
\affiliation{%
  \institution{University of Antwerp}
  \city{Antwerp}
  \country{Belgium}}
\email{joshua.krook@uantwerpen.be}

\author{Richard Hyde}
\affiliation{%
  \institution{University of Nottingham}
  \city{Nottingham}
  \country{UK}}
\email{richard.hyde@nottingham.ac.uk}

\author{Natalie Leesakul}
\affiliation{%
  \institution{University of Nottingham}
  \city{Nottingham}
  \country{UK}}
\email{natalie.leesakul@nottingham.ac.uk}

\author{Jeremie Clos}
\affiliation{%
  \institution{University of Nottingham}
  \city{Nottingham}
  \country{UK}}
\email{jeremie.clos@nottingham.ac.uk}

\author{Joel Fischer}
\affiliation{%
  \institution{University of Nottingham}
  \city{Nottingham}
  \country{UK}}
\email{joel.fischer@nottingham.ac.uk}

\renewcommand{\shortauthors}{Schneiders et al.}

\begin{abstract}
% MAX LENGTH: 150 WORDS, current 150
 Large Language Models (LLMs) are seemingly infiltrating every domain, and the legal context is no exception. In this paper, we present the results of three experiments (total \textit{N}~=~288) that investigated lay people's willingness to act upon, and their ability to discriminate between, LLM- and lawyer-generated legal advice. In Experiment 1, participants judged their willingness to act on legal advice when the source of the advice was either known or unknown. When the advice source was unknown, participants indicated that they were significantly more willing to act on the LLM-generated advice. \hl{The result of the source unknown condition was replicated in Experiment 2.} Intriguingly, despite participants indicating higher willingness to act on LLM-generated advice in Experiments 1 and 2, participants discriminated between the LLM- and lawyer-generated texts significantly above chance-level in Experiment 3. Lastly, we discuss potential explanations and risks of our findings, limitations and future work.%, and the importance of language complexity and real-world comparability.
 
 % Experiment 2, presents a direct replication, demonstrating the robustness of the findings. Experiment 3, investigates participants ability to identify the source of advice when it is unknown. Results of Experiment 1 show, replicated in Experiment 2, that participants to whom the advice source was unknown, express significantly higher willingness to act on the advice when it is generated by the LLM. Results of Experiment 3 shows that participants are significantly above chance in correctly distinguishing LLM- from lawyer-generated advice. Finally, this paper discusses limitations and future work, the importance of language complexity, real world comparability of the advice, as well as social acceptability biases.
\end{abstract}

\begin{CCSXML}
<ccs2012>
   <concept>
       <concept_id>10003120.10003121.10003122.10003334</concept_id>
       <concept_desc>Human-centered computing~User studies</concept_desc>
       <concept_significance>500</concept_significance>
       </concept>
   <concept>
       <concept_id>10003120.10003121.10003126</concept_id>
       <concept_desc>Human-centered computing~HCI theory, concepts and models</concept_desc>
       <concept_significance>300</concept_significance>
       </concept>
   <concept>
       <concept_id>10010405.10010455.10010458</concept_id>
       <concept_desc>Applied computing~Law</concept_desc>
       <concept_significance>300</concept_significance>
       </concept>
 </ccs2012>
\end{CCSXML}

\ccsdesc[500]{Human-centered computing~User studies}
\ccsdesc[300]{Human-centered computing~HCI theory, concepts and models}
\ccsdesc[300]{Applied computing~Law}
\keywords{Large language model, LLM, legal advice, generative AI, ChatGPT}
\begin{teaserfigure}
\centering
  \includegraphics[width=1\textwidth]{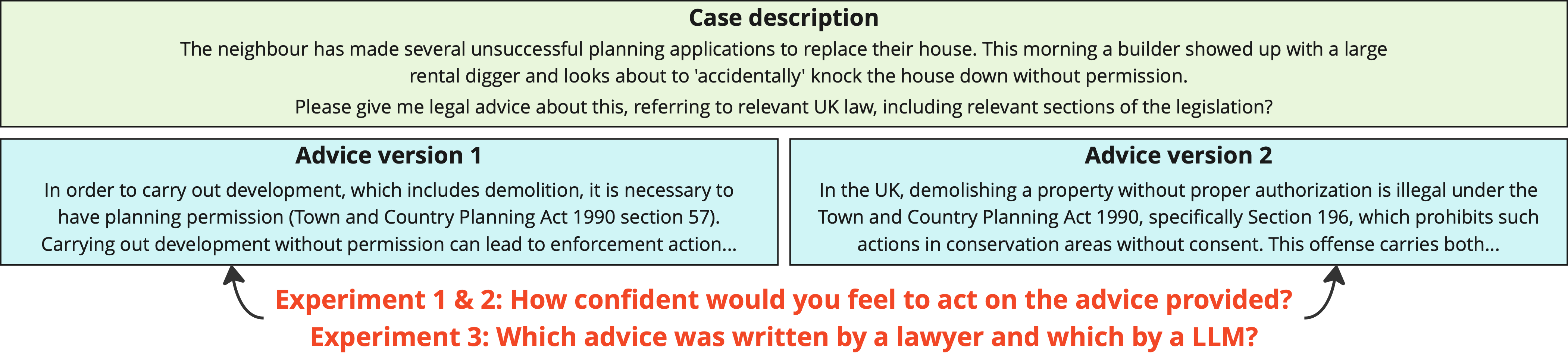}
  \caption{Legal advice provided by a lawyer (advice version 1) and a Large Language Model (LLM; advice version 2) in response to the same prompt. Experiments 1 and 2 evaluated participants' willingness to act on the legal advice, while Experiment 3 investigated participants' ability to discriminate between the advice sources.}
  % \textit{Both samples of advice have been shortened for this figure, full text can be found in} \href{https://osf.io/bksqa/?view_only=8c9a5893fb52478cb755870e56e686ca}{\texttt{Open Science Framework (OSF)}}
  %\caption{Participants flow in experiment 1. Following the pre-screening, participants received information on the study which was followed by the presentation of 18 legal cases (9 x lawyer generated advice and 9 x LLM generated advice). Participants were randomly allocated to the `Blinded' or `non-Blinded' condition, before ranking their willingness to act on each advice provided.}
  \label{fig:teaser}
\end{teaserfigure}

\maketitle

\input{Sections/Introduction}
\input{Sections/RelatedWork}
\input{Sections/Study}
%\input{Sections/Results}
\input{Sections/Discussion}

\input{Sections/Conclusion}

\begin{acks}
We would like to thank Horia Maior for his valuable comments on the manuscript. This project was supported by the Engineering and Physical Sciences Research Council [grant number EP/V00784X/1] UKRI Trustworthy Autonomous Systems Hub and Responsible AI UK [grant number EP/Y009800/1].\\

\textit{Data Access Statement:} For all experiments, the trial-level raw data and accompanying data analysis scripts are available on the Open Science Framework (see Section~\ref{sec:Transparency}).

\end{acks}

\bibliographystyle{ACM-Reference-Format}
\bibliography{biblio}
\newpage
\appendix
%%%% Commented out for accurate word count
\input{Sections/Appendix}
\end{document}

%% file: Sections/Introduction.tex
\section{Introduction}\label{sec:Introduction}
%GenAI, specifically LLMs, have received a lot of attention media and academia
%Furthermore, they have a high potential for impact on society at large
The emergence of \hl{Large Language Models (LLMS) as an example of generative artificial intelligence (GenAI)} 
%generative artificial intelligence (GenAI), particularly Large Language Models (LLMs), 
has shifted public attention towards the impact and tangibility of AI. OpenAI's ChatGPT, specifically, has captured the attention of the mainstream media, with many journalists focusing on its potential for  misuse\footnote{\textit{ChatGPT: US lawyer admits using AI for case research}: \url{https://www.bbc.co.uk/news/world-us-canada-65735769}}$^{,}$\footnote{\textit{Canada lawyer under fire for submitting fake cases created by AI chatbot}: \url{https://www.theguardian.com/world/2024/feb/29/canada-lawyer-chatgpt-fake-cases-ai}}. LLMs have also generated \hl{vast interest} in diverse research communities, including \hl{Artificial Intelligence (AI)}~\cite{zhao2024expel,zhu2024bootstrapping}, \hl{Human-Computer Interaction (HCI)}~\cite{Wang:2024:CHI,Weber:2024:CHI,Gero:2024:CHI}, and \hl{Computer-Supported Cooperative Work (CSCW)}~\cite{Wan:2024:CSCW,Karinshak:2024:CSCW,Wester:2024:CSCW}. For instance, 21\% of accepted full papers in the CHI '24 proceedings included the term `LLM' or `Large Language Model'\footnote{Full Query Syntax for the dl.acm.org: \texttt{ "query": \{ llm OR "large language model" \} "filter": \{ Conference Collections: CHI: Conference on Human Factors in Computing Systems, Book/Issue: CHI '24: Proceedings of the CHI Conference on Human Factors in Computing Systems, E-Publication Date: (01/01/2024 TO 12/31/2024) \}}}, demonstrating the clear interest in LLM research within the HCI community. 

With this surge in interest in LLMs, they are expected to continue to have wide-ranging impact on society. Examples include the healthcare sector, in which medical professionals have used LLM-based platforms to obtain fast summaries of patient health information~\cite{Yang:2024:LLMHealthcare} and in the financial sector, where prior research has shown shortcomings in relation to the ability of LLMs to reliably retrieve basic financial information~\cite{Lakkaraju:2023:LLMFinance}, \hl{and in the legal} context, \hl{where} the use and misuse of LLMs by legal experts has resulted in negative media attention~\cite{homoki2024large}\footnotemark[1]$^,$\footnotemark[2]. Similar observations have been reported with lay people, who reported high willingness to obtain legal advice from LLMs~\cite{Seabrooke:2024:TAS}. \hl{Given the ease of LLM use and the well documented risk of hallucinations---i.e., `textual output that is not consistent with [legal] facts'~{\cite{dahl2024large}}---the possible use of LLMs for high-risks context, such as providing legal advice, poses a potential risk to the general public. Therefore, understanding lay-people's willingness to act on the legal advice generated by LLMs is crucial, as this contributes to a deeper understanding that the impact that this `new' technology, and its wide ranging accessibility, has on society.}

%HCI, is an interdisciplinary discipline with origins in e.g., psychology,
%Therefore, we should learn from these disciplines and avoid making the same mistakes
%With the origins of Human-Computer Interaction (HCI) being in, amongst other, fields such as psychology~\cite{Carroll:1997:HCIasPsych}, we, as a research community, should strive to learn from our past, and avoid some of the same mistakes made. Dating as far back as 1968, the psychologist Jane Loevinger made the call to the American Psychologist Association (APA), calling for a greater emphasis on replicability of published experiments~\cite{ahlgren1969modest}. Around 40 years later, the field of psychology was in the midst of the `replication crisis', characterised by the failure to replicate existing results. In line with Loevinger's call for replication, the HCI community has over the last 15 years seen an increasing call for replicability through open science practices (e.g., open data and analysis scripts) and a greater emphasis on replications of experiments, thereby preventing a replication crisis within HCI~\cite{Wilson:2012:Replication,Hornbak:2014:Replication,Echtler:2018:Replication,Feger:2019:Replication,Cockburn:2020:Replication}.\\
%\hl{Note:}\url{https://www.slideshare.net/slideshow/replichi-graduate-student-perspectives/7972980#4}\url{https://www.cs.nott.ac.uk/~pszmw/pubs/RepliCHI-panel\_CR.pdf}\url{https://research.google/pubs/replichi-sig-from-a-panel-to-a-new-submission-venue-for-replication/}\url{https://dl.acm.org/doi/pdf/10.1145/2468356.2479636}

%This paper presents...
In this paper, we present three experiments investigating lay peoples perceptions of LLMs in the context of legal advice. In doing so, we investigate two research questions:
\vspace{1mm}
\begin{enumerate}
    \item Are lay people more willing to act on LLM- or lawyer-generated legal advice?
    \item Are lay people, when unaware of the source, able to identify the source of legal advice presented to them?
\end{enumerate}
\vspace{1mm}
Building on existing literature~\cite{Seabrooke:2024:TAS}, Experiment 1 (\textit{N}~=~100) investigated whether lay people report higher willingness to act upon LLM- or lawyer-generated legal advice\hl{, depending on whether the source of the advice was known or unknown}. Experiment 2 (\textit{N}~=~78) replicate\hl{d} the key manipulation of Experiment 1\hl{, i.e., the source unknown condition.} Lastly, Experiment 3 (\textit{N}~=~110), investigated participants ability to discriminate between LLM- and lawyer-generated legal advice when the source was unknown to them. Experiment 1, replicated in Experiment 2, showed that participants reported significantly higher willingness to act on LLM-generated advice than lawyer-generated advice, when the advice source was unknown to them. %, express significantly higher willingness to act on LLM-generated advice. 
In Experiment 3, participants showed above chance level when discriminating between LLM- and lawyer-generated advice, when the source was unknown.

%We conclude the paper by discussing the implications as well as future directions informed by the here presented findings.

We conclude the paper by discussing the importance of language complexity, the risks and strategies to mitigate risks that overtrust on LLMs can bring, as well as limitations and future work.

\subsection{Openness and Transparency Statement}\label{sec:Transparency}
Experiment 1 was not preregistered, but Experiments 2 and 3 were. For all experiments, the trial-level raw data and accompanying data analysis scripts are available on the \href{https://osf.io/bksqa/?view_only=8c9a5893fb52478cb755870e56e686ca}{Open Science Framework (OSF)}\footnote{\url{https://osf.io/bksqa/?view_only=8c9a5893fb52478cb755870e56e686ca}}. We report the rationale for the sample sizes, all manipulations, measures, and data exclusions. All three experiments were approved by the Institutional Review Board (IRB) of the University of Nottingham (CS-2023-R22).

%% file: Sections/RelatedWork.tex
\section{Related Work}\label{sec:RelatedWork}
This section highlights relevant literature on the impact LLMs have had on the legal landscape to date. The section also notes additional high risk domains, beyond the legal context, that have been affected by LLMs. Lastly, we present relevant literature on trust and general perceptions towards AI.

\subsection{LLMs in the Legal Landscape}
LLMs are increasingly being used by legal professionals for a range of tasks, from legal research and e-discovery, to contract drafting and the filing of motions in court~\cite{aoshearman2023harvey,homoki2024large}. In many cases, law firms are limiting the use of LLMs to summarising information, digesting large numbers of documents (in discovery, for example), and providing summaries for lawyers to review~\cite{kapoor2024promisespitfallsartificialintelligence}. 
Litigation matters involving large companies can involve tens of thousands of documents, meaning that LLMs serve a practical and functional use-case in these scenarios.

On the other hand, some lawyers have already been caught-out relying on LLMs (such as ChatGPT and Gemini) to generate legal advice for clients, often running into the problem of hallucinated facts~\hl{{\cite{dahl2024large}}}. In Mata v. Avianca~\cite{Mata:Avianca}, for example, a lawyer relied on falsely generated cases while seeking a damages claim against an airline on behalf of his injured client. %\hl{add reference to Mata v Avianca - https://law.justia.com/cases/federal/district-courts/new-york/nysdce/1:2022cv01461/575368/54/} 
The court rebuked the lawyer, stating that the provision of false or misleading cases wastes the time of the court and the opposing counsel. This misuse of the court's time interrupts the court's schedule, potentially impacting other cases, and wasting public money~\cite{Harber}. %(\hl{Cite: Harber v Commissioners for His Majesty’s Revenue and Customs, 2023).}

Courts in the United States have thus far taken a dim view of the use of LLMs for legal filings, noting the risk of errors and hallucinations and the necessary due diligence of legal professionals. In another case involving Google's Bard (now Gemini), a lawyer was again found to provide false cases to the court, which then had to try to find the non-existent cases~\cite{cohen2024opinion}. %As the problems of LLMs become more widely known, courts may start issuing harsher reprimands \hl{such as being referred to the regulatory body for further assessment of consequences going beyond fining the lawyer}~\cite{HUMPHREYS}. 
\hl{As the problems with LLMs become more widely recognised, courts may impose stricter reprimands, such as the referral to regulatory bodies for further evaluation and consequences beyond simply fining the lawyer}~\cite{HUMPHREYS}. This is in part due to the serious nature of lying to the court, something taken seriously in past legal precedent~\cite{Villiers}.% \hl{Cite: (Villiers v Villiers, 2021).}

There are nevertheless circumstances where an LLM could provide high level of accuracy, while still falling below the threshold of an expert's legal advice. \citet{nay2023large} investigated the use of an LLM, specifically models developed by OpenAI, in the context of legal analysis for tax law. %, or better phrased as \textit{tax law advisers of attorneys}. 
Their findings demonstrated that \textit{``LLMs, particularly when combined with prompting enhancements and correct legal texts, can perform at high levels of accuracy but not yet at expert tax lawyer levels.''} This suggests that LLMs, even with prompt enhancements, are not yet able to reach professional performance in the domain of, e.g., tax law. %a human-in-the-loop to review the output%, to reach the same level of expertise as a professional tax lawyer.
%https://dl.acm.org/doi/10.1145/3630106.3659048
\hl{To investigate when and how LLM could be useful in the legal context, Cheong et al.~{\cite{LLM:2024:Facct}} conducted seven online workshops with legal professionals. Participants were presented with legal cases and, through a series of workshop activities, identified 25 dimensions informing safe and effective LLM usage for the legal context. Cheong et al.~{\cite{LLM:2024:Facct}} suggest a series of guiding principles informing legal advice provided by LLMs. Specifically, the authors suggest that LLMs refrain from providing actual context specific advice but limit their suggestions to `non-opinion facts' about, e.g., specific laws.}
% \subsection{LLMs in the legal context}
% \hl{Note:} This section will look at prior literature investigating the legal domain and its relationship to LLMs. This needs a particular focus on HCI research (and CHI), e.g.,~\cite{LLM:2024:Facct,Seabrooke:2024:TAS,Jiang:2023:ICAIL,Weber:2024:CHI}

% Link between Information Retrieval, LLM, and Law. When a court would like to search for previously judged cases, Information Retreaval Combined with LLM can help trim examples of legal cases from the past. Yixiao Ma et al. present a Legal Case Retrieval Dataset for Chinese Law System (LeCaRD). 

% LLMs have the ability to synthesise large amounts of documents and information; This has significant implications in the legal domain, and Yang et al. \cite{yang2024large} stress the fact that LLMs such as ChatGPT have the potential to transform various aspects of legal practice and access to justice.

% Earlier works presented in 2018 CAIL2018: A Large-Scale Legal Dataset for Judgment Prediction \cite{xiao2018cail2018} which included the largest legal dataset. 

% \cite{homoki2024large} discuss the limitations and customisation requirements of LLMs, particularly for legal uses, and show how they have already shown great performance in essential legal tasks such as text retrieval, generation, labelling, and classification.

% LLM in Human Rights and Criminal Law \cite{jimma2022college}
LLMs are influencing the legal landscape not only from a professional perspective, but also in ways that might have impact on non-legal experts~\cite{Jiang:2024:LeveragingLLMs,Seabrooke:2024:TAS}. \citet{Seabrooke:2024:TAS} %conduct a survey based study (\textit{N}~=~105) in which they 
investigated lay people's  willingness to obtain legal advice in a wide variety of domains (e.g., traffic, divorce, planning, property, or civil disputes). % that while only a minority of participants report having first hand experience with LLMs for the legal context (17\%), 
Nearly half of the participants (45\%) said that they would be likely to use LLMs to inquire legal advice in the future. Furthermore, they highlight that the willingness to generate legal advice is not evenly distributed, and varies greatly depending on the specific domain (e.g., civil disputes (25\%) vs. tenancy law (58\%)). However, while these results suggest that lay people are willing to generate advice using LLMs, the authors did not investigate whether lay people are willing to \textit{act} on it. 

% \hl{Note:} \citet{Jiang:2024:LeveragingLLMs} show that LLMs can be used to educate non-legal experts on legal concepts by utilising storytelling as a method to increase relatability, and thereby knowledge retention, to participants own lives.

\subsection{LLMs in other Domains}
Large language models have been deployed and studied in numerous domains \hl{in which non-experts interact, or rely on advice provided, by LLMs. Examples include} healthcare~\cite{nadarzynski2019acceptability,miles2021health,reis2024influence,Krook:2023:TASHubResponse}, journalism, story telling, communication, and public messaging~\cite{waddell2018robot,asscher2023human,chi2024effect,papachristos2024adapting}, and education~\cite{liu2024beyond,bernabei2023students,steenstra2024engaging,tayan2023considerations,lim2023generative}. In healthcare, \hl{numerous studies have investigated how non-medical experts interact and accept medical advice provided by LLM chatbots.} \citet{nadarzynski2019acceptability}, for instance, examined the acceptability of AI-led chatbot services and found that while most participants were open to using chatbots for healthcare, concerns about accuracy and lack of human touch were barriers to adoption. \citet{miles2021health} further explored this issue by investigating how perceived stigma and severity of health issues influence acceptance of chatbots, finding that while chatbots might be suitable for sensitive health issues due to increased anonymity, they are less preferred for severe health conditions. \citet{ayers2023comparing} compared chatbot responses to those of physicians on a public social media forum, demonstrating that chatbot-generated responses were often preferred for their quality and empathy. Finally, \citet{reis2024influence} investigated the public perception of AI-generated medical advice, highlighting an 'anti-AI bias' in which advice, even when supervised by physicians, was deemed less reliable and empathetic compared to solely human-generated advice. \hl{Riedl et al.~{\cite{riedl2024patients}} presented similar results using a large scale within-subject (N~=~1183) study, demonstrating that participants preferred a human doctor, followed by an AI-assisted human doctor, and lastly the AI system by itself.}

\citet{waddell2018robot} examined how audiences perceive news articles attributed to machine authors and found that machine-authored news is often deemed less credible due to lower perceived anthropomorphism and unmet expectations. \citet{asscher2023human} investigated human evaluations of machine translation, particularly in ethically sensitive situations (e.g., communication between users with power imbalance), and found a negative bias against machine-translated content even among professional translators.

The effectiveness of LLMs in public messaging has also been a topic of interest. \citet{chi2024effect} explored the impact of AI chatbots on individuals' attitudes towards environmental protection and their willingness to pay for conservation efforts. The study found that the problem solving capabilities of chatbots were particularly influential in encouraging financial contributions. \citet{oviedo2023risks} illustrated risks of using LLMs for safety-related advice, revealing that they often provide over-simplified, erroneous, or biased information.

In educational settings, LLMs have become an object of interest both for their potential as a tool for personalisation of learning and for their risk to become a tool for undetectable cheating. \citet{liu2024beyond} showed the potential of using an LLM as a virtual teaching assistant, enhancing student engagement and providing personalised feedback, while also highlighting the need for human supervision. \citet{bernabei2023students} examined ChatGPT's use in engineering education, revealing its ability to improve understanding and expedite assignments, but also emphasising the importance of critical evaluation and ethical considerations. \citet{steenstra2024engaging} demonstrated the potential of LLMs in helping the development of health education games, while underscoring the need for human oversight for the accuracy of the content. \citet{tayan2023considerations} focused on the impact of ChatGPT on technology courses, advocating its use as a complementary tool, and providing guidelines for responsible implementation. \citet{lim2023generative} offered a paradoxical perspective on Generative AI in education, emphasising the need to embrace its potential as a tool to support learners while addressing academic and ethical challenges that stem from its growing capabilities as a generator of content.

\subsection{Trust, Perception, and Acceptability of AI}\label{sec:RW_TrustPerceptionAcceptAI}
\hl{An increasing amount of research has emphasised the growing impact AI-generated content has on user trust}~\cite{jakesch2019ai,Liu:2022:AIConcole,Mink:2022:DeepPhish}\hl{, user perception}~\cite{logg2019algorithm,castelo2021conservatism,bigman2018people}\hl{, and ultimately user acceptance of AI~{{\cite{wester2024exploring}}.}}

The perception of AI influences how people feel about its outputs. \citet{jakesch2019ai} identified the `Replicant Effect,' where individuals exhibit distrust toward Airbnb hosts believed to have AI-generated profiles, particularly in mixed environments with both AI and human-authored profiles. Although AI aversion is prevalent, \citet{logg2019algorithm} observed that individuals often prefer algorithmic over human judgement in specific contexts. Their research revealed that people tend to adhere more to advice when they believe it originates from an algorithm rather than a person, showcasing algorithm appreciation. This preference was consistent in various domains, including numerical estimations, forecasting tasks, and even subjective judgements such as romantic attraction. \hl{Not just the context of the advice~{\cite{logg2019algorithm}}, but also the characteristics of the human receiver play a role with respect to how LLM-generated advice is received~{\cite{wester2024exploring}}. The increase of} use of LLMs for advice-seeking has prompted investigations into user characteristics and perceptions. \citet{wester2024exploring} explored how people perceive LLM-generated advice and the influence of user characteristics on these perceptions. Their findings suggest that personality traits, particularly agreeability and technological insecurity, play a role in shaping how people evaluate and engage with LLM advice.

%trust in AI
\hl{Liu et al.~{\cite{Liu:2022:AIConcole}} investigated trust in emails allegedly written with complete human agency, shared human \& AI agency, or complete AI agency across three contexts of increasing interpersonal emphasis: product inquiry (low), party invitation (mid), and pet loss consolation (high). In a between-subject design, participants rated the same emails with varying supposed authorship. The authors measured trustworthiness using the three dimensions: ability, benevolence, and integrity~{\cite{Mayer1995integrative}}. Results showed that higher AI agency reduced perceived trustworthiness, with an increase in AI agency being linked to a reduction of trustworthiness. Furthermore, trustworthiness was affected by its degree of interpersonal emphasis, with higher emphasis resulting in higher trust ratings.}

Acceptance of AI is not uniform and appears to be influenced by various factors. For example, \citet{castelo2021conservatism}  found that political conservatism correlated with AI aversion, particularly for consequential tasks. This aversion stems from conservatives' increased risk perception associated with AI adoption. Furthermore, \citet{bigman2018people} discovered a general aversion to machines making moral decisions, mainly due to the perception that machines lack a complete mind capable of both thinking and feeling. The perceived lack of agency and experience in machines contributes to this reluctance, suggesting that the nature of the task and the perceived capabilities of AI play a role in shaping people's acceptance of AI in decision-making roles.

%https://www.usenix.org/conference/usenixsecurity22/presentation/mink
\hl{A recent study by Mink et al.~{\cite{Mink:2022:DeepPhish}} investigated how people perceive artificially generated social media profiles (LinkedIn), depending on if they were real or AI-generated. Specifically, the authors manipulated the profile picture as well as the `about 'description of the person. Furthermore, the authors present three different types of prompts (no prompt, soft prompt, hard prompt) to investigate its impact on the perceived trustworthiness. The authors find that both the artifacts, i.e., AI-generated content, and the prompting significantly decrease the trustworthiness and acceptance of the AI profiles.}

%\subsection{Summary}
%As the here presented research shows, LLMs have already been extensively used by professionals in the legal context. However, perhaps just as concerning is the willingness by lay-people to use LLMs in a variety of context, including high risk ones such as: legal advice, healthcare or education. In this paper we focus on the legal domain as an example of a high-risk context with potentially severe consequences. 

%% file: Sections/Study.tex
\input{Sections/Exp1}
\input{Sections/Exp2}

\input{Sections/Exp3}

%% file: Sections/Exp1.tex
% \section{Study Design}\label{sec:Study}
%This paper presents three studies. In study 1 we investigate non-lawyers willingness to act on legal advice generated by a lawyer or LLM (N=100). Study 2 is a replication study (N=50), verifying the robustness of the findings presented in the first experiment. Lastly, study 3 assesses participants ability to distinguish LLM from lawyer generated advice when being blinded to the source. All presented studies received ethical approval by the Institutional Review Board (IRB) of the University of \texttt{Blinded for Review}.% (CS-2023-R22).

%\input{Sections/ExamplePrompts}

\section{Experiment 1}\label{sec:Study1}
In Experiment 1, we investigated participants' willingness to act on legal advice generated by an LLM or a lawyer, both when the source of advice was known and unknown. \hl{We showed all participants a series of legal cases and asked them} to rate their willingness to act on the legal advice using a scale from 1 (`Strongly Disagree') to 9 (`Strongly Agree'). \hl{We explicitly informed half of the participants} of the source of each advice provided, being either LLM- or lawyer-generated, \hl{while we did not inform the remaining participants}. The key questions were whether participants would be more willing to act on the legal advice provided by a LLM or by a lawyer, and whether this would interact with whether the source of the advice was known to participants or not.

\subsection{Method}
\subsubsection{Participants}
When considering the number of participants required for Experiment 1, we did not have a clear estimate of the effect size for the interaction between the advice source (LLM vs. lawyer) and group (source known vs. source unknown) factors. We therefore recruited 100 participants, with 50 participants in each group. A power analysis in G*Power~\cite{faul2007PowerAnalysis} estimated that 98 participants would provide 95\% power to obtain a small-to-medium effect (\textit{n} $\approx$ 98, \textit{f}~=~0.15, 1-$\beta$~=~.95, $\alpha$~=~.05), using the default values of 0.5 for the correlation between measures and 1 for the nonsphericity correction.

For all three experiments, participants were recruited from Prolific and we excluded any participants who failed at least two out of three attention checks. \hl{The first attention check, presented during the demographic questions, asked participants to press a highlighted key on their keyboard (`w'). The second and third attention check simply asked to select a specific number on the Likert scale presented (`4' and `7').} No participants were excluded in Experiment 1. As the legal prompts and responses were based on UK law, we only recruited participants who self reported fluency in English, were between 18--60 years old, and were currently living in the UK. 

The final, overall sample for Experiment 1 consisted of 100 participants (60 male, 40 female, 0 prefer not to say) who were aged between 19 and 58 years (\textit{M}~=~33.26 years, \textit{SD}~=~9.00 years). The source unknown group consisted of 50 participants (29 male, 21 female, 0 prefer to not to say), who were aged between 19 and 58 years (\textit{M}~=~32.92 years, \textit{SD}~=~9.06 years). The source known group consisted of 50 participants (31 male, 19 female, 0 prefer not to say), who were aged between 19 and 53 years (\textit{M}~=~33.60 years, \textit{SD}~=~9.01 years).

%\hl{@Eike, deal with the demographic for Set1 and Set2, see comment below.}
%overall 60 male, 40 female, min age 19, max age 58, mean age 33.26, sd 9.00
%blinded 29 male, 21 female, min age 19, max age 58, mean age 32.92, sd 9.06
%nonblinded 31 male, 19 female, min age 19, max age 53, mean age 33.60, sd 9.01
%The chi-square statistic for gender is 0.1667. The p-value is .683091. The result is not significant at p < .05.

The median completion time was 14:53 minutes but, as progression in the experiment was self-paced, the completion time varied between participants. Participants were compensated with £9/hour. %Participants were compensated with 9£/h and had a median completion time of 14:32 minutes.

\input{Sections/Prompt_1}

%Participants were compensated with 9£/h and had a median completion time of 14:53 minutes.

\subsubsection{Prompts}\label{sec:prompts}
%\hl{Do we want to bring the Lix number / word count discussion up front?}
Based on prior literature~\cite{Seabrooke:2024:TAS}, we selected three areas within the legal context for which participants reported a high likelihood to use LLMs for the generation of legal advice, specifically: traffic law, planning law, and property law. In consultation with legal professionals, we identified six prompts for each of these three areas, resulting in a total of 18 prompts. These prompts were inspired by real questions asked online in the subreddit \href{https://www.reddit.com/r/LegalAdviceUK/?rdt=38701}{\url{r/LegalAdviceUK}}\footnote{\url{www.reddit.com/r/LegalAdviceUK}}\hl{~{\cite{LLM:2024:Facct}}}.
All prompts were based on UK law. Example prompts as well as LLM- and lawyer-generated answers for each category can be seen in~\Cref{fig:PlanningLawExample,fig:PropertyLawExample,fig:TrafficLawExample}. All 18 prompts and the corresponding LLM- and lawyer-generated answers are available in the OSF (see~\Cref{sec:Transparency}). \hl{We used ChatGPT-4o to generate the LLM-generated prompts}, while \hl{UK-based lawyers generated the lawyer-generated advice. The recruited lawyers had expertise} within the three domains: traffic, planning, and property. \hl{Two of the three lawyers, specialising in property and planning, were recruited using the personal networks of the authors. The last lawyer, specialising in traffic law, was one of the authors of the paper. However, all three lawyers were unaware of the LLM-generated responses to the same prompts, nor were they privy to the nature of the study design in which their prompts would be used. They were} asked to generate succinct advice based on the provided prompts. The lawyer-generated advice was often preceded with a re-iteration of the case details, which was removed to ensure comparability in brevity between the LLM and lawyer-generated advice. We verified, with legal experts, that no text was removed which changed the meaning of the advice. No text was added to the advice provided. %All advice generated, both by the LLM and the lawyers, was comparable in length. 
Each prompt was shown the same number of times in each experiment, and all three experiments used the same prompts and responses. \hl{All three lawyers were unpaid and provided their services in-kind.}

%\subsubsection{Measurements}
%Previous research has demonstrated that participants are willing to use LLM to \textit{enquire} legal advice~\cite{Seabrooke:2024:TAS}. However, it has yet to be determined to what extend the LLM generated advice meets non-legal experts quality requirements to an extend that they are \textit{willing to act} on the advice provided.

\subsubsection{Experimental Design and Measurements}
Experiment 1 followed a 2 (advice source: LLM vs. lawyer) $\times$ 2 (group: source known vs. source unknown) mixed factorial design. Advice source and group were manipulated within-subjects and between-subjects, respectively. 
The dependent variable was participants' ratings of their willingness to act on each legal advice provided, which were ranked on scale from 1 (`Strongly Disagree') to 9 (`Strongly Agree').
All three experiments were implemented using jsPsych~7\footnote{https://www.jspsych.org/latest/} and hosted using JATOS\footnote{https://www.jatos.org}.

\input{Sections/Prompt_2}

\subsubsection{Procedure}
\hl{We presented participants with} the participant information sheet highlighting relevant information including details on the data being collected, participants' right to withdraw, and information on the procedure of the experiment. Once participants consented to their participation, we collected demographic data on age, gender, and language fluency. As part of the demographic data, we had a simple attention check (\textit{`Please press the highlighted key on your keyboard'}). \hl{Subsequently, we informed participants} that some of the legal advice they would be presented with had been generated by lawyers, while other advice had been generated by an LLM. To ensure a common understanding of what an LLM was, a short, easily understandable, description was provided (see~\Cref{app:LLM}). 

Participants were randomly allocated to the source known or unknown group, ensuring that we had an equal number of participants in each group. Each participant was presented with 18 legal cases. These cases were written by either an LLM (9) or a lawyer (9) using the same 18 prompts. The LLM and the lawyer answered the same short legal prompts within each of the three categories: traffic law, planning law, and property law~\cite{Seabrooke:2024:TAS}. Two additional attention check cases were presented at random times, resulting in a total of 20 trials. In the source known condition, participants saw the statement:  `Case \#: This advice has been written by [an LLM / a lawyer]' above the prompt and the advice. In the source unknown condition, by contrast, each prompt and advice was simply placed below the statement:  `Case \#', with the author of the advice omitted. For each case, participants were asked to answer \textit{`In the described situation, I would feel confident acting on the advice provided'} on a 9-point  scale from 1 (`Strongly Disagree') to 9 (`Strongly Agree'). Participants had to respond to each case in order to move on with the task, and progression throughout the task was self-paced.

% \begin{figure}
%     \centering
%     \includegraphics[width=.6\linewidth]{Figures/REGALS_Exp1_Setup.png}
%     \caption{Participants distribution}
%     \label{fig:enter-label}
% \end{figure}

% \begin{table}[]
% \begin{tabular}{@{}lll@{}}
% \toprule
%  & Lawyer & LLM \\ \midrule
% Set 1 cases & 4, 5, 6 & 1, 2, 3 \\
% Set 2 cases & 1, 2, 3 & 4, 5, 6 \\ \bottomrule
% \end{tabular}
% \caption{Case distribution for Set 1 and Set 2. This distribution ensures that the same prompt is never shown twice, once with lawyer advice and once with LLM advice, to the same participants}
% \label{tab:CaseDistri}
% \end{table}

%Experiment 2, in order to assess the potential harm caused by the potential reliance on LLM generated advice, investigates the quality of the LLM generated advice as well as lawyers ability to distinguish the two sources when blinded.
%Experiment 2, a within subject design using lawyers as participants, uses \textsc{Source} (\textit{lawyer} or \textit{LLM}) as within subject manipulation while, contrasting experiment 1, all participants are blinded to the source of the advice. Following each presented prompt and advice, participants are asked to rank the \textsc{Quality of Advice} using a \hl{9}-point Likert scale (`Very low quality' to `Very high quality'), as well as their \textsc{Confidence in Source} (1: `Definitely LLM generated' to 9: `Definitely lawyer generated').

\subsection{Experiment 1: Results}

\subsubsection{Data Analysis}
All analyses were performed in RStudio~\cite{R:2023}. We report both frequentist analyses and Bayes factors to provide estimates of the evidence in favour of the alternative (H1) and null (H0) hypotheses. The evidence categories proposed by~\citet{jeffreys1961theory}, and the corresponding interpretations proposed by~\citet{Lee:Wagenmakers:2014}, are shown in~\Cref{table:BF} and were used to interpret the Bayes factors. All Bayes factors were calculated using the default priors provided in the `BayesFactor'~\cite{Morey:Rouder} package (version 0.9.12.4.7). Estimated Bayes factor values are reported with the error percentage of each estimate. 

\begin{table}[h]
\begin{tabularx}{\linewidth}{sb}
%\begin{tabularx}{@{}ll@{}}
\toprule
\textit{BF}$_{10}$ & Interpretation \\ \midrule
> 100 & Extreme evidence for H1 \\
30 - 100 & Very strong evidence for H1 \\
10 - 30 & Strong evidence for H1 \\
3 - 10 & Moderate evidence for H1 \\
1 - 3 & Anecdotal evidence for H1 \\
1 & No evidence \\
0.33 - 1 & Anecdotal evidence for H0 \\
0.10 - 0.33 & Moderate evidence for H0 \\
0.03 - 0.10 & Strong evidence for H0 \\
0.01 - 0.03 & Very strong evidence for H0 \\
< 0.01 & Extreme evidence for H0 \\ \bottomrule
\end{tabularx}
\caption{Bayes factor evidence categories according to~\citet{jeffreys1961theory} and corresponding interpretations by~\citet{Lee:Wagenmakers:2014}.}
\label{table:BF}
\end{table}

\begin{figure}
    \centering
    \includegraphics[width=\linewidth]{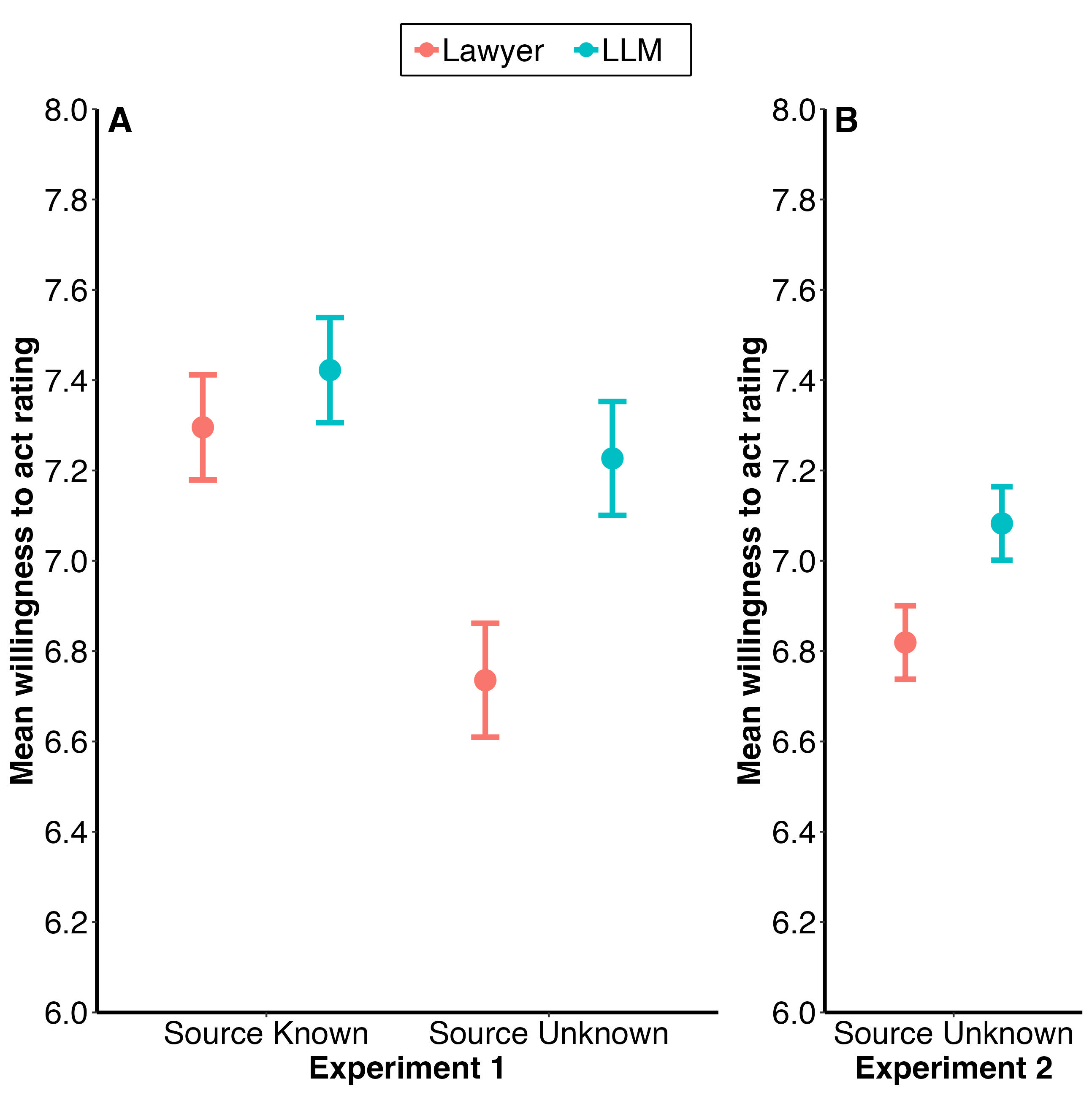}
    \caption{Mean willingness to act ratings in Experiments 1 and 2. Error bars represent difference-adjusted, within-subjects, 95\% confidence intervals~\cite{Baguley:2012:CalculatingANOVA}.}
    \label{fig:Data_Exp1and2}
\end{figure}

\input{Sections/Prompt_3}

\subsubsection{Willingness to Act on Legal Advice}
Participants' mean ratings for their willingness to act on the legal advice, separated by advice source and experimental group, are shown in Panel A of~\Cref{fig:Data_Exp1and2}. \hl{Levene's test indicated that the assumption of homogeneity of variance was not violated, \textit{F}(1,~98)~=~1.69,~\textit{p}~=~.20. }A 2 (advice source: LLM vs. lawyer) $\times$ 2 (group: source known vs. source unknown) mixed analysis of variance (ANOVA) on participants' mean ratings revealed a significant main effect of source, \textit{F}(1, 98)~=~13.09, \textit{p}~<~.001, generalised eta squared ($\eta_{\text{G}}^{2}$)~=~.02. The corresponding Bayes factor provided very strong evidence for the alternative hypothesis, \textit{BF}$_{10}$ $\approx$ 40.27 ($\pm$ 1.96\%). Participants gave significantly higher ratings to the LLM-generated advice (\textit{M}~=~7.32, \textit{SD}~=~1.09) than the lawyer-generated advice (\textit{M}~=~7.02, \textit{SD}~=~1.11). No significant main effect of group was observed, with the Bayes factor analysis providing anecdotal support for the alternative hypothesis, \textit{F}(1, 98)~=~3.58, \textit{p}~=~.06, $\eta_{\text{G}}^{2}$~=~.03, \textit{BF}$_{10}$~$\approx$~1.25~($\pm$~1.51\%). Collapsed across LLM- and lawyer-generated advice, the source unknown group (\textit{M}~=~6.98, \textit{SD}~=~1.21) gave comparable ratings to the source known group (\textit{M}~=~7.36, \textit{SD}~=~0.96). Finally, and most interestingly, there was a significant interaction between the advice source and group factors, \textit{F}(1, 98)~=~4.56, \textit{p}~=~.04, $\eta_{\text{G}}^{2}$~=~.01. The Bayes factor provided anecdotal evidence for the alternative hypothesis, \textit{BF}$_{10}$~$\approx$~1.58~($\pm$~2.47\%).

Following the significant advice source $\times$ group interaction, paired \textit{t}-tests were conducted to compare the effect of advice source in each group. In the source unknown group, participants rated their willingness to act on the LLM-generated advice (\textit{M}~=~7.23, \textit{SD}~=~1.17) significantly more highly than the lawyer-generated advice (\textit{M}~=~6.74, \textit{SD}~=~1.21), \textit{t}(49)~=~3.91, \textit{p} < .001, Cohen's \textit{d$_z$}~=~0.55. Moreover, the Bayes factor provided very strong evidence in favour of the alternative hypothesis, \textit{BF}$_{10}$~=~89.70 ($\pm$ 0\%). The source known group, by contrast, showed no significant difference in ratings for advice that was generated by the LLM (\textit{M}~=~7.42, \textit{SD}~=~1.00) and the lawyers (\textit{M}~=~7.30, \textit{SD}~=~0.93), \textit{t}(49)~=~1.09, \textit{p}~=~.28, \textit{d$_z$}~=~0.15. The corresponding Bayes factor provided moderate support for the null hypothesis, \textit{BF}$_{10}$~$\approx$~0.27 ($\pm$ 0.05\%).

%% file: Sections/Prompt_1.tex
\begin{figure}[ht]
%\raggedright
\begin{tcolorbox}[colback=white, colframe=black, title=Example: Traffic Law 4]
{\normalsize
\textbf{Prompt:}
    %\textit{``My son was caught driving without a license, despite having earned his license last year. When he was pulled over by the patrol, he sped off rather than answering their questions. How much trouble is he in? Can you give me some advice about his legal situation, considering relevant UK laws and sections of legislation?''}\\

\textit{``My father got a ticket saying he broke Section 5(1)(a) of the Road Traffic Offenses law. What happened was he rode over a speed bump and got stuck in a carriageway, preventing cars from going past him. An angry driver came over to him and started banging on his windscreen, threatening to call the police. My father ran away from this angry man before the police arrived. Can you give me some advice about my legal situation, considering relevant UK laws?''}\\
\vspace{3mm}
    \textbf{Advice provided by LLM:}\\
    \textit{``Under Section 5(1)(a) of the Road Traffic Offenses law in the UK, it is an offense to use a vehicle in a manner that causes unnecessary obstruction of the road. In your father's case, being stuck on a speed bump and blocking the carriageway could be seen as such an obstruction. However, it is also important to note that he fled the scene due to fear for his safety from the aggressive driver. This context could be relevant in his defense, as it demonstrates a potential threat that prompted his departure. It is advisable to gather any available evidence, such as witness statements or CCTV footage, that supports his account of the events, and to consult with a solicitor who can help argue that his actions were a response to imminent danger rather than a willful obstruction.''}\\
    %\textit{``Your son is in significant legal trouble for driving without a license, which he had earned but apparently did not possess at the time, and for evading police, which can be charged under the Road Traffic Act 1988. Driving without a license can lead to a fine up to £1,000 and penalty points, while failing to stop for police and dangerous driving (Section 2 of the Road Traffic Act 1988) can result in severe penalties, including an unlimited fine, driving ban, and up to 2 years in prison. It's crucial to engage a solicitor specializing in motoring offences to mitigate the consequences, gather evidence, and ensure compliance with all legal requirements.''}\\
\vspace{3mm}
    \textbf{Advice provided by lawyer:}\\
    \textit{``If your father was driving a car (which seems to be implied by the banging on the windscreen, but you say that he “rode” over the speed bump), I presume that your father has received a notice saying that he acted contrary to section 5 of the Road Traffic Regulation Act 1984. This prohibits acting contrary to a Traffic Regulation Order which is in force for a particular road. Road Traffic Regulation Orders can prohibit or restrict the waiting of vehicles in particular areas. If there was a Road Traffic Regulation Order in place it may be that when your father ran away from the angry driver he was seen to have breached this prohibition. Acting contrary to a Road Traffic Regulation Order is an offence which can lead to a fine of up to £1000. However, the circumstances you describe (if they can be proved, for example through dashcam footage) may amount to mitigation, and lead to a much lower penalty. Your father should consult a solicitor as soon as possible.''}
    %\textit{``It is an offence to fail to stop when required to do so by the police (Road Traffic Act 1988 section 163). The police must be in uniform and they must have been making the request in the course of their duty. This offence can be tried in the magistrates court and lead to an unlimited fine. It is not an offence to drive in the UK without a copy of a licence. If you do not have your licence when requested, you have one week to produce it at the appropriate police station (Road Traffic Act 1988 section 164). Provided that the licence that he gained last year was still valid (he has not had it taken away due to a conviction for a driving offence or for accumulation of points) there would have been no offence committed if he had been unable to produce his licence on request. I would advise your son to consult a solicitor as soon as possible.''}
}
\end{tcolorbox}
\caption{\hl{Example of prompt for Traffic Law (\#4) with corresponding LLM and Lawyer generated advice.}}
\label{fig:TrafficLawExample}
\end{figure}

%% file: Sections/Prompt_2.tex
\begin{figure}
\begin{tcolorbox}[colback=white, colframe=black, title=Example: Property Law 2]
{\normalsize
\begin{quote}
    \textbf{Prompt:}\\
    \textit{``My landlord has threatened to evict me from my apartment because I have a dog. They have now told me there is a "no pets" clause in the contract, but promised me it was okay to have a dog when I first moved in. Could you advise me about the legality of this in UK law, with reference to relevant legislation?''}\\

    \textbf{Advice provided by LLM:}\\
    \textit{``In the UK, the legality of pet ownership in rental properties hinges on the terms of the tenancy agreement and relevant housing legislation. While landlords can include "no pets" clauses, previous verbal consent for pet ownership could constitute a waiver or variation of contractual terms, provided there is evidence to support this claim. The Housing Act 1988 allows for such clauses, but landlords must also act reasonably and fairly towards tenants, and breaches of the Consumer Rights Act 2015 may also be relevant. If negotiations fail, seeking legal advice or assistance from housing charities like Shelter is advisable, as the outcome depends on the specifics of the agreement, relevant laws, and potential legal recourse.''}\\

    \textbf{Advice provided by lawyer:}\\
    \textit{``On first impressions there are two legal avenues we could consider. The first is that the terms of your lease were formally varied by the conversation with your landlord. This would require the law to accept that a lease can be varied orally, in contrast to the usual practice that dealings with land need to be made in writing. This is a hard argument to make since longer residential leases tend to be varied via a Deed of Variation. What is more, if your lease agreement contains a provision that the terms can only be varied in writing, then the recent Supreme Court ruling in Rock Advertising Limited confirms the binding effect of such clauses. Nevertheless, if such a clause is absent, an argument could be made that it is possible to vary orally. For instance, the requirement for writing under the Law of Property Act 1925 relates to creating or disposing of an interest in land, but here the interest in land (the lease) has already been given to you in writing and what we would be arguing is that this is a subsequent change involving no such creation or disposal. As a result the variation can be binding even if agreed orally.''}
\end{quote}
}
\end{tcolorbox}
\caption{Example of prompt for Property Law (\#2) with corresponding LLM and Lawyer generated advice.}
\label{fig:PropertyLawExample}
\end{figure}

%% file: Sections/Prompt_3.tex
\begin{figure}
\begin{tcolorbox}[colback=white, colframe=black, title=Example: Planning Law 6]
{\normalsize
\begin{quote}
    \textbf{Prompt:}\\
    \textit{``I am in the middle of building a fence with my neighbour's property, 
    who has objected to the proposed colour of the fence. The council has 
    rejected my planning permission and told me that I cannot use a colour and 
    must install a basic timber fence, and that our planned metal fence does 
    not comply with local regulations. Please give me legal advice about this, 
    referring to relevant sections of UK law?''}\\

    \textbf{Advice provided by LLM:}\\
    \textit{``In the context of UK property law, it's essential to adhere to local 
    planning regulations enforced by the council, as failure to do so can 
    result in legal consequences. The rejection of planning permission and 
    insistence on installing a basic timber fence instead of a proposed 
    metal one signifies non-compliance with local regulations. Sections 106 
    and 14 of the Town and Country Planning Act 1990 and the Planning 
    (Listed Buildings and Conservation Areas) Act 1990 empower local 
    authorities to enforce conditions to ensure compliance with policies, 
    particularly in conservation areas. To avoid potential fines and 
    enforcement actions, it's advisable to comply with the council's 
    directives, seek guidance on alternative options, and consult a 
    legal professional specializing in property law and planning regulations 
    for further assistance.''}\\

    \textbf{Advice provided by lawyer:}\\
    \textit{``You do not need to obtain planning permission to erect a fence provided that your proposed fence complies with the requirements in the Town and Country Planning (General Permitted Development) Order 2015 Schedule 2 Part 2 Paragraph A. The requirements relate to the height of the fence, and do not limit the material or colour of the fence. However, such limitations may be imposed if your property is situated in a national park or a conservation area, or if your property is a listed building. If this is the case, you need to ensure that a proposed fence meets the requirements of the conservation area or national park authority, and/or that you obtain listed building consent before erecting your fence. Failure to do this may mean that enforcement action can be taken against you, which may involve you being required to remove the fence. Failure to comply with such a requirement may amount to a criminal offence.''}
\end{quote}
}
\end{tcolorbox}
\caption{Example of prompt for Planning Law (\#6) with corresponding LLM and Lawyer generated advice.}
\label{fig:PlanningLawExample}
\end{figure}

%% file: Sections/Exp2.tex
\section{Experiment 2}

With HCI originating, in part, from psychology  ~\cite{Carroll:1997:HCIasPsych,Dix:2017:HCIPsych}, HCI research would benefit from the open science and research integrity practices that psychology has adopted in recent decades. Dating as far back as 1968, psychologist Jane Loevinger urged the American Psychological Association to place greater emphasis on replicability of experimental research~\cite{ahlgren1969modest}. Around 40 years later, psychology found itself in the midst of a replication crisis. Mirroring Loevinger's thoughts, over the last 15 years the HCI community has also seen an increasing call for replicability through open science practices and a greater emphasis on replications~\cite{Wilson:2012:Replication,Hornbak:2014:Replication,Echtler:2018:Replication,Feger:2019:Replication,Cockburn:2020:Replication}.

In Experiment 2, we therefore sought to replicate the key result from Experiment 1. Specifically, we aimed to replicate the finding that participants reported that they were more willing to act on LLM-generated legal advice than advice that had been generated by lawyers. To this end, we re-ran the source unknown condition from Experiment 1 with a new set of participants. 

% discussion of importance of replication here.

\subsection{Method}

\subsubsection{Participants}
In Experiment 1, the effect size for the source manipulation in the source unknown group was \textit{d}$_z$~=~0.55. A power analysis in G*Power~\cite{faul2007PowerAnalysis} indicated that 45 participants would provide 95\% power to replicate an effect of this size (\textit{n} $\approx$ 45, \textit{d}$_z$~=~0.55, 1-$\beta$~=~.95, $\alpha$~=~.05). We preregistered that we would therefore over-sample slightly and recruit 50 participants. However, upon recruiting a random set of 50 participants from Prolific, we found that the gender distribution was highly imbalanced. We therefore recruited the minimum number of additional participants (\textit{n}~=~29) required to balance the gender distribution as closely as possible, while taking care to match the number of participants (and gender) within each counterbalancing condition. The recruitment of these additional participants was not preregistered.

After recruiting the additional 29 participants, one participant was excluded because they failed two out of three attention checks (as per our preregistered exclusion criteria). The final sample consisted of 78 participants (39 female, 38 male, 1 prefer not to say), with 39 participants in each counterbalancing condition. The participants were aged between 18 and 60 years (\textit{M}~=~37.86 years, \textit{SD}~=~11.01 years) and all reported that they spoke fluent English. The experiment was not advertised to participants who had participated in Experiment 1.

As Experiment 1, progression in Experiment 2 was self-paced and the completion time varied between participants. The median completion time was 14:32 minutes. Participants were compensated with £9/hour. %Participants were compensated with 9£/h and had a median completion time of 14:32 minutes.

\subsubsection{Experimental Design and Measurements}
The experiment followed a repeated-measures design with one independent variable: advice source (LLM vs. lawyer). As in Experiment 1, the dependent variable was participants' rating of their willingness to act on each legal advice shown.

\subsubsection{Prompts and Procedure}
The prompts and procedure were identical to those of Experiment 1, except that only the source unknown condition from Experiment 1 was included. %The experiment lasted approximately 12 minutes but, as progression in the experiment was self-paced, the completion time varied between participants. Participants were compensated with £9/hour.

\subsection{Experiment 2: Results}
Participants' mean willingness to act ratings are shown in Panel B of~\Cref{fig:Data_Exp1and2}. A paired sample \textit{t}-test demonstrated that participants gave significantly higher ratings to the texts that were generated by an LLM (\textit{M}~=~7.08, \textit{SD}~=~1.17) than those generated by lawyers (\textit{M}~=~6.82, \textit{SD}~=~0.99), \textit{t}(77)~=~3.22, \textit{p}~=~.002. The effect size was small-to-medium, \textit{d}$_z$~=~0.37, and the Bayes factor indicated that there was strong evidence for the alternative hypothesis, \textit{BF}$_{10}$~=~14.00. This finding replicates the key result from Experiment 1, thereby demonstrating that it is robust and replicable.

%% file: Sections/Exp3.tex
\section{Experiment 3}
Previous research (see~\Cref{sec:RW_TrustPerceptionAcceptAI}\hl{)}, e.g.,~\cite{logg2019algorithm,chi2024effect} has demonstrated that participants, at times, tend to be more prone to adhere to algorithmic advice over human generated advice. Our findings from Experiment 1 and 2 align with these findings, showing that participants also report higher willingness to act on LLM-generated legal advice \hl{when the source of advice is unknown.}

In Experiment 1 and 2, participants rated their willingness to act on the advice provided differently depending on whether the source of the advice was known or not. \hl{Given this finding, and the known risk of LLMs to produce inaccurate content, i.e., hallucinations, a question arises: were participants able to discriminate between the advice generated by LLMs and that by of the lawyers?} Experiment 3 addressed this question by investigating participants' ability to identify the source of \hl{legal advice when it was not disclosed to them}. Participants were presented with the same 18 legal cases and advice texts from Experiments 1 and 2, and were asked to rate the extent to which they thought each advice text had been generated by an LLM or lawyer. %Participants are asked to rank their confidence in the source being either a LLM or a lawyer on a 6-point Likert scale.

\subsection{Method}
\subsubsection{Participants}
As per our preregistration, we recruited 110 participants (55 male, 53 female, 2 prefer not to say) between 18 and 57 years (\textit{M}~=~35.20 years, \textit{SD}~=~9.59 years) using Prolific. All participants reported that they spoke fluent English. We applied the same inclusion/exclusion criteria as in the previous experiments, but additionally excluded all participants who had participated in either Experiment 1 or 2. As in the previous two experiments, participants had to answer two of the three attention check questions correctly in order to be included in the final dataset. We did not exclude any participants due to failed attention check questions. 
%Consistent with Experiment 1 \& 2, this experiment was advertised with a completion time of approximately 12 minutes. Participants were compensated with £9/hour.

Consistent with Experiments 1 and 2, progression in Experiment 3 was self-paced and the completion time varied between participants. The median completion time was 12:12 minutes and participants were compensated with £9/hour. %Participants were compensated with 9£/h and had a median completion time of 14:32 minutes. 

\subsubsection{Experimental Design and Measurements}
Experiment 3 followed a repeated-measures design with one independent variable: advice source (LLM vs. lawyer). The dependent variable was participants' rating of their confidence in whether the presented advice was generated by an LLM or a lawyer for each advice presented. For each advice text, participants were asked to indicate their confidence in the source by answering \textit{`Please rate to what extent you think the text has been generated by an LLM or a lawyer'} on a 6-point scale from 1 (`Definitely LLM generated') to 6 (`Definitely lawyer generated').

\subsubsection{Prompts and Procedure}
The prompts were identical to those of the first two experiments. However, as Experiment 3 aimed to identify participants' ability to identify the source of the advice when it was unknown, we made slight changes to three LLMs responses: `Planning advice 2', `Traffic advice 3', and `Property advice 4'. Specifically, we removed the three/four first words stating `As a lawyer,...' or `As your legal advisor,...'. As in Experiment 2, we only included the source unknown condition. Apart from the change in question and the rating scale, see above, Experiment 3 followed the same procedure as Experiment 1 and 2.

%Following each advice participants were asked to indicate their confidence in the source of advice using 6-point Likert scale%`Please rate to what extent you think the text has been generated by a LLM or a lawyer' on a 6-point Likert scale (`Definitely LLM generated' to `Definitely lawyer generated'). 

\subsection{Experiment 3: Results}
As per our preregistration, we used a measure derived from signal detection theory to estimate participants' ability to discriminate between the LLM- and lawyer-generated advice~\cite{macmillan2002signal}. Specifically, we calculated the area under the receiver operating characteristic (ROC) curve (AUC). Unlike mean ratings~\cite{modirrousta2023effects, modirrousta2023gamified}, and mean rating difference scores~\cite{HIGHAM2024101785}, ROC analysis allows discrimination to be measured separately from response bias. In the current context, discrimination refers to participants' ability to accurately distinguish between the LLM- and lawyer-generated texts. Response bias, by contrast, refers to participants' overall tendency to rate all texts as LLM- or lawyer-generated. We focus on AUC as a bias-free measure of participants' discrimination performance, although note that the ROC curve also provides a visual representation of response bias.

ROC analysis can be employed wherever participants are tasked with discriminating between two categories using graded discrimination ratings on a multi-point scale. In our discrimination task, participants made such graded ratings by judging the extent to which the advice was generated by an LLM or a lawyer on a scale from 1 (``Definitely LLM generated'') to 6 (``Definitely lawyer generated''). Therefore, given the upper boundary of the scale refers to lawyer-generated advice, it is helpful to conceptualise the task as one requiring participants to detect lawyer-generated advice. Under this definition, trials in which lawyer-generated advice is presented can be considered `signal' trials, whereas trials that contain LLM-generated advice can be considered `noise' trials. When making judgements, participants are assumed to base their responses on a `decision variable' that is determined by their subjective evaluation of the amount of signal in the text~\cite{stanislaw1999calculation}. They are also assumed to adopt a decision threshold, which is known as a `criterion'. If the decision variable is sufficiently high that it meets or exceeds the decision threshold, the participant will give a positive response, thereby classifying the trial as a signal trial. Traditionally, signal trials (i.e., lawyer advice trials in the current context) in which a participant correctly concludes that a signal is present are considered `hits'. Conversely, noise trials (i.e, LLM advice trials) in which a participant incorrectly concludes that a signal is present are considered `false alarms'.

To measure discrimination, each point on the rating scale is treated as a separate threshold reflecting different LLM/lawyer criteria. For each participant, the proportion of hits and false alarms (i.e., the hit and false alarm rates, HRs and FARs) is then calculated for each point on the rating scale. For example, for scale point 2, the HR and FAR would be defined as the proportion of lawyer- and LLM-generated advice that received a rating of 2--6, respectively. The HR and FAR rate for each scale point can be plotted against each other to calculate an ROC curve, which can be used to visualise both discrimination and response bias.

\begin{figure}
    \centering
    \includegraphics[width=.9\linewidth]{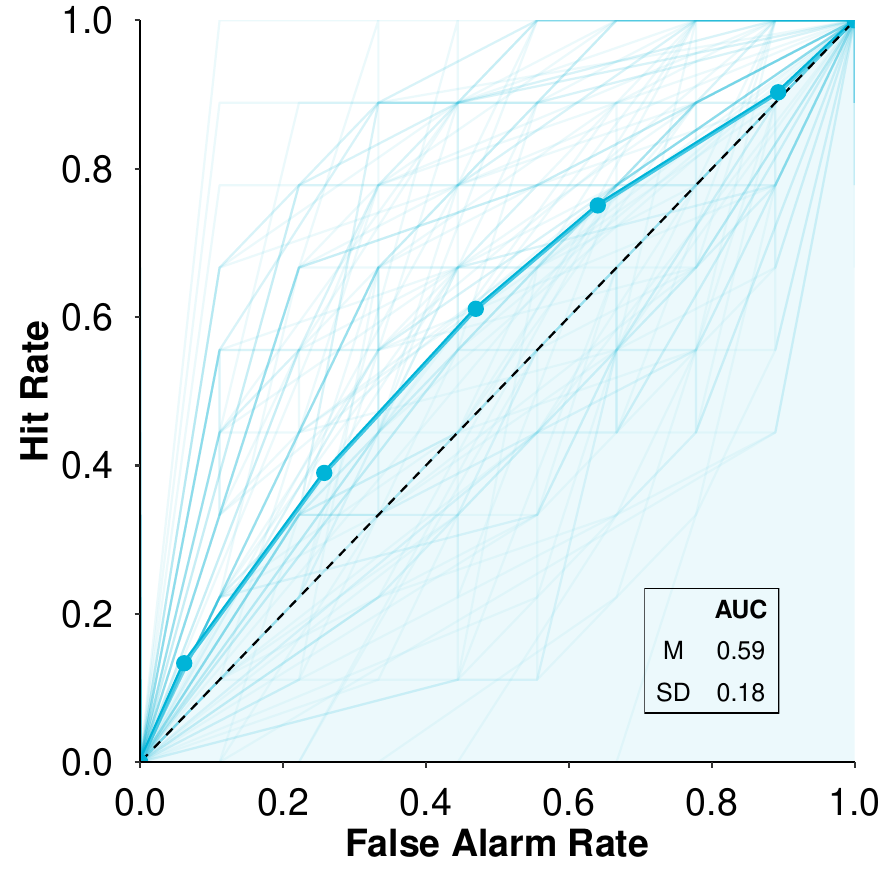}
    \caption{ROC curve for Experiment 3 indicating, through the bowing towards the top-left corner, above chance discrimination of LLM- and lawyer-generated legal advice. The area under the ROC curve (AUC) is highlighted in light blue (\textit{M}~=~.59, \textit{SD}~=~.18). Faint lines represent individual participant ROC curves.}
    \label{fig:ROC}
\end{figure}

The ROC curve for Experiment 3, with the mean HRs and FARs plotted against each other, is presented in blue in~\Cref{fig:ROC}. The points on the ROC curve reflects the different scale values, with scale point 1 depicted on the [1,1] coordinates. The diagonal, dashed line that runs from the [0,0] to [1,1] coordinates is a reference line that represents chance discrimination performance. The more the ROC curve bows away from the diagonal line, towards the top-left corner of the plot, the better the discrimination. The position of the scale points also provides a visual indication of response bias. Specifically, points clustering towards the bottom-left corner of the plot indicate a conservative response bias, i.e., participants tending to rate the presented advice as LLM-generated, whereas points clustering towards the top-right of the plot indicate a liberal response bias, with participants tending to rate the presented advice as lawyer-generated.

As noted above, we quantified discrimination by estimating the AUC, which we calculated using the trapezoidal rule~\cite{pollack1969sampling, higham2019new}. AUC values vary between 0 and 1, with .50 representing chance discrimination performance (i.e, the diagonal, dashed line in~\Cref{fig:ROC}) and 1 representing perfect discrimination. Following our preregistration, we compared participants' AUC values to a theoretical mean of .50 to establish whether participants could discriminate between the lawyer- and LLM-generated advice significantly above chance. This analysis confirmed that participants could discriminate the source of the advice \hl{with an AUC of 0.59 (SD~=~.18; or a 59\% probability of correctly ranking lawyer-generated advice above LLM-generated advice)}. The Bayes factor provided extreme evidence in favour of the alternative hypothesis, \textit{t}(109)~=~5.51, \textit{p}~<~.001, \textit{d}~=~0.53, \textit{BF}$_{10}$~=~5.10 $\times$ 10$^{4}$.

%% file: Sections/Discussion.tex
% \hl{Not sure yet where this goes}
% As we showed in Experiment 1 and 2, our participants exhibited algorithm appreciation~\cite{logg2019algorithm}, i.e., the tendency to prefer or rely on algorithms over humans, even when blinded to the source. Similar observations have been made in numerous domains~\citet{ayers2023comparing}. \citet{wester2024exploring}

% However, on the contrary algorithm aversion is equally distributed, e.g.~\citet{waddell2018robot}
% \citet{reis2024influence}
\section{Discussion}\label{sec:Discussion}
In this paper, we have presented three experiments. Experiment 1 investigated lay people's willingness to act on LLM- and lawyer-generated legal advice. The results showed that, when the source of the advice was unknown to participants, they were more willing to act on the LLM-generated advice than the lawyer-generated advice. This result was replicated in Experiment 2. In Experiment 3, we investigated lay people's ability to distinguish between the LLM- and lawyer-generated advice when the source was unknown. Participants discriminated between the advice significantly above chance, but their discrimination performance was far from perfect.

The remainder of the discussion focuses on the importance of the phrasing of the language used and the risks of overtrust in LLM-generated content and strategies to mitigate this. Lastly, we will present limitations while also integrating future research directions at the intersection of LLMs and the legal domain throughout.

%\hl{The remainder of the discussion focuses on the importance of the language complexity and advice length of the legal advice provided, furthermore, we will discuss the degree of real world comparability of the advice provided in the here presented experiments. Lastly, this section will present the limitations of the current study design, while also integrating future research directions at the intersection of LLMs and the legal domain throughout.}

\subsection{Use of Language for Legal Advice}
In Experiments 1 and 2, participants reported a significantly higher willingness to act on the legal advice provided by the LLM than the lawyers when the source was unknown. While the presented work here cannot answer \textit{why?} with certainty---and this is therefore a challenge for future work---we now discuss potential explanations for our results.

\subsubsection{Language Complexity and Advice Length.} 
While the LLM- and lawyer-generated advice was broadly comparable in length, we chose not to equate the word counts or textual complexity exactly. Indeed, the average number of words used in the lawyer-generated advice was 170 (min: 107; max: 276), while the average number of words used by the LLM-generated advice was 124 (min: 93; max: 176). %, \textit{t}(17)~=~4.01, \textit{p}~<~=.0005, showing that the LLM-generated advice was significantly shorter.
% to avoid artificially restricting the expressions of either the LLM or the lawyers. %As adjusting the length might be easier for the LLM then for the lawyer, a solution for this would be to generate the LLM advice using the lawyers word count as an instruction for the LLM prompt. 
We also did not restrict the complexity of the language that the lawyers or the LLM used. The LLM-generated advice appeared to be more complex, as measured by an average Lix score~\cite{Bjornsson:1983:LIX} of 72 (min: 60; max: 79) versus 57 (min: 49; max: 65) for the lawyer-generated advice\footnote{Lix ranking: Very Easy: 20; Easy: 30; Average: 40; Difficult: 50; Very Difficult: 60.}. These Lix scores correspond to `Very Difficult' and `Difficult' to comprehend for the LLM- and the lawyer-generated advice respectively~\citep[Table 4]{Bjornsson:1983:LIX}. %that the complexity of the LLM-generated advice was `Very Difficult' to comprehend~\citep[Table 4]{Bjornsson:1983:LIX} while the lawyer-generated advice was classified as `Difficult'~\citep[Table 4]{Bjornsson:1983:LIX} to comprehend. 
While balancing both of these metrics could have been done, and would have increased the advice similarity between sources, we were concerned that doing so would reduce realism and impose experimenter bias on the advice. Limiting the LLM or the lawyer to an arbitrary language complexity or word count would introduce constraints that would distort the advice that both an LLM and lawyer would otherwise generate. Therefore, while such experimenter-imposed constraints increase experimental control, they also reduce realism and ecological validity.

One possible explanation, is that participants were more willing to act on the LLM-generated advice as they conflated complexity with quality of advice. To test this possibility, future work could investigate people's willingness to act on legal advice, when the complexity is matched between the LLM and the lawyers. %The public perception of lawyers might also have a role here, as the public might expect a particular level of complexity from lawyers due to the popular conception of lawyer jargon, the use of Latin, and the official nature or gravitas of court environments.

\subsubsection{Comparability to Real World Legal Advice}
Our study directly compared participants' responses to LLM- and lawyer-generated advice. For practical purposes, we stipulated that the reply to our prompt had to be a short `summary' of the legal advice to be provided to a potential client. In practice, a lawyer would rarely, if ever, provide such a short one-paragraph summary to a client, nor would an LLM without this added stipulation. %Traditional legal advice provided by a lawyer is often several pages long, composing long emails, attached documents, correspondence, and so on, and therefore more readily identifiable as `human' than in the context of our study. 

Lawyers do not typically provide one-paragraph answers to a legal problem, nor are they trained to do so in an expeditious manner. While the LLM could provide an answer all-but instantaneously, the lawyers often took days or weeks to provide legal advice following our queries. There are practical reasons for this difference. A lawyer must worry about their professional reputation, their licensing certificate, and the risk of being sued for negligence, when advising clients regarding legal matters. By nature, these precautions do not lend themselves to one-paragraph answers. Instead, lawyers are often prone to hedge their bets, providing both sides of an argument, and cautioning against certainty and finality when providing solutions to legal problems.

LLMs have the opposite problem, as they often appear overly confident when providing advice. Such overconfidence---which might be a consequence of the vocabulary, sentence structure, and syntax used---can lead to clients being exposed to decisively-voiced, hallucinated~\hl{{\cite{dahl2024large}}} facts. %This confidence is not an emotion in the context of an LLM, but a result of the vocabulary used, sentence structure and syntax. 
In part, LLMs are trained on the rules of grammar, to present convincing, structured responses with clear introductions, arguments, conclusions, sign-posting, and other features of good writing. These structural advantages cannot easily be compared to a human writer responding off-the-cuff to a prompt. To attain the same level of writing structure and quality, a human writer may need to edit, or have their work edited, something which was not done in the context of our experiments.

%\subsection{Causes for Higher Ratings and Overtrust}
\subsection{\texorpdfstring{\hl{Higher Ratings, Overtrust, and LLM Characteristics}}{High Ratings, Overtrust, and LLM Characteristics}}
Our findings of highlight (Experiment 1 and 2) that participants, while being able to distinguish the source above chance (Experiment 3), have a higher willingness to act on the legal advice when generated by an LLM. In this section we discuss possible causes for this, risks that might result of this overtrust, as well as possible strategies addressing the overtrust in LLM-generated legal advice.

% \hl{As our findings highlight, it seems that the willingness to act or not does not solely depend on the source of advice, but participants might rely on different heuristics. This section discusses participants reliance on other indicators beyond the source as well as the risks which might come from an overtrust on AI-generated advice.}

\subsubsection{Ratings are Influenced by more than just the Source}
When examining our findings, a contradictory observation can be made. In Experiment 1 and 2, participants in the source unknown condition, reported a higher willingness to act on LLM-generated advice than on the lawyer-generated advice. On the other hand, Experiment 3 showed that when participants were directly asked to identify the source, they were able to distinguish between LLM- and lawyer-generated advice at an above chance rate. Although these findings were observed in separate experiments, when considered together, they suggest that participants might have been aware of the source in Experiment 1 and 2 even when it was not disclosed to them. Interestingly, while participants seemed to be---at least to some extent---aware of the source, even when it was not provided to them, they behaved differently than in the source known condition of Experiment 1. Specifically, participants increased their willingness to act ratings when explicitly informed that the advice presented had been generated by a lawyer. 
% When looking closer at Experiment 1, we can observe that the mean willingness to act ratings for the LLM-generated texts only differ slightly between the source unknown and the source known conditions (source unknown: \textit{M}~=~7.23, \textit{SD}~=~1.17; source known: \textit{M}~=~7.42, \textit{SD}~=~1.00). In contrast, the mean willingness to act ratings for the lawyer-generated advice shows significant differences between the the two conditions (source unknown: \textit{M}~=~6.74, \textit{SD}~=~1.21; source known: \textit{M}~=~7.30, \textit{SD}~=~0.93). This suggests that participants, when being aware of the source of the advice, did not lower their ratings for the LLM, but increased their rating for the lawyer-generated advice.
This pattern may indicate a social acceptability bias~\cite{Zhang:2024:CHI} in which participants thought that they \textit{should} be more willing to act on lawyer-generated advice. Participants may have given higher ratings to the lawyer-generated advice than the LLM-generated advice in the source known condition, as this aligns with perceived social norms that humans should be trusted over an algorithm. %Therefore, differences in ratings for the source known condition are non-significant.

% This pattern might indicate, that participants did not purely rely on their perception of the sources quality to assess their willingness to act on the advice presented. Instead, a different heuristic might be used by participants. In the source known condition the difference in rating between the two different advice sources, might be affected by social acceptability bias~\cite{Zhang:2024:CHI}. Participants might increase their rating for the human-generated advice, over the AI-generated advice, as this aligns with social acceptable norms to trust a lawyers advice over the advice provided by an algorithm. Therefore, differences in ratings for the source known condition are non-significant.

% Furthermore, as presented in~\Cref{sec:RW_TrustPerceptionAcceptAI}, prior research~\cite{wester2024exploring,Volkel:2021:CUI} has identified that not just the advice plays a key role in how users perceive and accept it, but the users---and their individual characteristics---themselves. \citet{wester2024exploring} show that participants with high `agreeableness' level are more likely to receive LLM generated advice positively. While the authors demonstrate this for general questions (e.g., related to weight loss advice), the identification of `who', i.e., which user groups, is receptive to LLM-generated advice is highly relevant in high risk scenarios such as the legal context. Therefore, future work would benefit from an investigation of if specific user groups, e.g., defined by users personality traits, are particularly receptive towards advice in high risk domains.

\subsubsection{Risks of Overtrust}
Overtrust in LLM-generated advice carries an abundance of risks, some of which are being specifically addressed through policies and regulations such as the European Union Artificial Intelligence Act (EU AI Act). For instance, the EU AI Act - article 50.2\footnote{Article 50: Transparency Obligations for Providers and Deployers of Certain AI Systems} emphasises that \textit{``Providers of AI systems, including general-purpose AI systems, generating synthetic audio, image, video or text content, shall ensure that the outputs of the AI system are marked in a machine-readable format and detectable as artificially generated or manipulated...''}~\cite{EU_AI_Act_Article50}. While these regulations aim to ensure the safe use of AI-based systems, including LLMs, there has been a growing focus on how to effectively implement them. One strategy directly addressing the above article is the implementation of watermarks~\cite{Kirchenbauer:2023:Watermark,Tang:2024:LLMDetection}. However, while watermarks may allow machines to detect AI-generated content, they may not (necessarily) improve the transparency of AI-generated text for people. Although watermarks enable the generation of \textit{``artificially generated or manipulated text in a machine-readable format...''}~\cite{EU_AI_Act_Article50} they are still \textit{``invisible to humans''}~\cite{Kirchenbauer:2023:Watermark}. 

While machine detectable indicators of AI-generated content are highly valuable to enable automatic detection, improving the general public's AI literacy will become increasingly important as AI-infused systems become more sophisticated and prevalent. In Experiment 3, participants were significantly above chance when discriminating LLM- from lawyer-generated advice, but there was still clear room for improvement \hl{(59\% probability of ranking the laywer-generated advice higher than the LLM-generated advice)}. %However, as AI-generated content continues to influence an increasing number of domains, the (human) 
% equipping humans with the skills necessary to distinguish between artificial and human-generated content becomes increasingly relevant. 
Within the related field of fake news detection, current research~\cite{modirrousta2024inductive} has demonstrated that interventions---in the form of short interactive training session---can significantly improve participants' discrimination of true and fake news headlines. We suggest that future work draws inspiration from the fake news detection literature, in order to develop similar interventions that improve the general public's AI literacy, and equips lay people with the skills to discriminate human- from AI-generated content. %, ultimately improving their ability to discriminate artificial- from human-generated content. 
Such interventions may be particularly useful for user groups who are especially susceptible to trusting LLM-generated content, such as people with high `agreeableness'~\cite{wester2024exploring}.%, as this might put them at additional risk of relying on hallucinations~\cite{wester2024exploring}.

% As presented in~\Cref{sec:RW_TrustPerceptionAcceptAI} prior research~\cite{wester2024exploring,Volkel:2021:CUI} has identified that not just the advice plays a key role in how users perceive and accept it, but the users---and their individual characteristics---themselves. \citet{wester2024exploring} show that participants with high `agreeableness' level are more likely to receive LLM generated advice positively. While the authors demonstrate this for general questions (e.g., related to weight loss advice), the identification of `who', i.e., which user groups, is receptive to LLM-generated advice is highly relevant in high risk scenarios such as the legal context. Therefore, future work would benefit from an investigation of if specific user groups, e.g., defined by users personality traits, are particularly receptive towards advice in high risk domains.

Beyond watermarking and human interventions, the EU AI Act contains further provisions (Article 13)\footnote{Article 13: Transparency and Provision of Information to Deployers} for `high-risk' systems, including transparency around the disclosure of risks, limitations of the systems, accuracy level, and impacts on health, safety, and fundamental rights. These `high risk' provisions may also apply to AI systems offering legal services. The obligation on providers of these systems to offer further information on risks and disclaimers could address some of the problems of watermarking. However, the extent to which these safeguards are adopted in AI-based systems varies greatly. For instance, the current disclaimer provided by ChatGPT-4o simply states that \textit{``ChatGPT can make mistakes. Check important info.''} while Google's Gemini states that \textit{``Gemini may display inaccurate info, including about people, so double-check its responses''}\footnote{Both disclaimers were observed on the 20.08.2024 and might change in the future.}. In addition to the brevity of the disclaimers,  their usefulness depends on users reading, understanding, and acting upon the information provided. %For example, if a user is told that a chatbot can provide misleading advice, the question remains whether they will read this disclaimer, and, more importantly, consider the potential risks before deciding to act on the legal advice provided. 
Prior work has shown that users rarely read or engage with online terms and conditions containing legal information~\cite{Robinson:2020:TermsAndService}. Therefore, informing users of potential risks of the use of LLMs may not be sufficient.
% A presumption that transparency will always lead to rational consumer behaviour is not necessarily borne out by this related literature.

% - not the quality of advice, but other heuristics, for instance ease of processing the advice?

% \hl{Exp 1/2: The salience of the advice's' source is not obvious as participants haven't explicitly asked to identify it. Therefore, while they are aware that some are LLM while other are lawyer generated, they might not engage in deeper processing on the origin of the advice, as they do in Experiment 3.}

% \subsection{`Who' Acts on LLM-generated Legal Advice?}
% As presented in~\Cref{sec:RW_TrustPerceptionAcceptAI} prior research~\cite{wester2024exploring,Volkel:2021:CUI} has identified that not just the advice plays a key role in how users perceive and accept it, but the users---and their individual characteristics---themselves. \citet{wester2024exploring} show that participants with high `agreeableness' level are more likely to receive LLM generated advice positively. While the authors demonstrate this for general questions (e.g., related to weight loss advice), the identification of `who', i.e., which user groups, is receptive to LLM-generated advice is highly relevant in high risk scenarios such as the legal context. Therefore, future work would benefit from an investigation of if specific user groups, e.g., defined by users personality traits, are particularly receptive towards advice in high risk domains.

\subsubsection{\texorpdfstring{\hl{LLMs: Competencies and Identifiability}}{LLMs: Competencies and Identifiability}}
\hl{As discussed in Section~{\ref{sec:RelatedWork}}, prior research (e.g.,}~\cite{LLM:2024:Facct,nay2023large}) \hl{has shown that, when used appropriately, LLMs can offer value in legal settings. Cheong et al.~{\cite{LLM:2024:Facct}}, in line with the EU AI Act, highlighted a significant concern regarding the accuracy of LLM-generated legal advice as part of their `AI Capability’ dimension. Furthermore, they emphasised that the law is not static and the legal field is constantly evolving. This raises questions such as, `Are LLMs equipped to consider recent legal changes when providing advice?' and `Do they possess the contextual understanding needed, including relevant facts, geographic location, and applicable laws?' Unlike legal professionals, LLMs lack an obligation to adhere to the justice system, nor are they being held accountable. As demonstrated in this paper, participants were able to distinguish between LLM- and lawyer-generated advice above chance-level, but not greatly above-chance (59\% probability of correctly ranking lawyer-generated advice above LLM-generated advice). While the EU AI Act, as described above, argue for AI-generated content to be detectable by AI, Cheong et al.{~\cite{LLM:2024:Facct}} instead proposed that LLMs should not `pretend to be human', i.e., humans should be able to easily identify LLM-generated content. In the same way as Article 13 of the AI Act, this would strive to increase transparency, and ultimately reduce the risk of over-reliance and unrealistic expectations. To investigate to what extent the LLM-generated advice presented in this paper adheres to the recommendations set in {\cite{LLM:2024:Facct}} remains a task for future work.}% Adhering to this recommendation would, presumably, improve lay-peoples ability to discern LLM-generated from lawyer-generated advice. To verify to what extent the LLM-generated advice presented in this paper adheres to the recommendations set in {\cite{LLM:2024:Facct}} remains a task for future work.}

%\subsection{Limitations}
\subsection{\texorpdfstring{\hl{Limitations and Future Work}}{Limitations and Future Work}}
To compile the lawyer-generated advice, we recruited three specialists lawyers within the domains of traffic, planning, and property law. These lawyers likely have a personal writing style that may not transfer to other lawyers. \hl{Furthermore, as described in}~\Cref{sec:prompts}, \hl{we removed text preceding the actual core advice provide. This was only necessary for the property law as the lawyer generating advice provided verbose advice. As described, we verified with legal experts that the condensed advice still kept the original meaning. Nevertheless, the changing of the original text might have impacted the results.} Likewise, we used ChatGPT-4o to create the LLM-generated advice. Different results might be achieved with other LLMs. Therefore, future work should test the boundary conditions of our findings with different advice sources.

Furthermore, as our focus was on lay peoples' perceptions of LLM- and lawyer-generated advice, we did not evaluate the accuracy of the advice. Given that we were interested in participants' willingness to act on and discriminate the LLM- and the lawyer-generated advice, the accuracy of the advice is inconsequential for this purpose. Nevertheless, future work would benefit from establishing the validity of the advice---particularly that of the LLM---as LLMs are prone to create hallucinations~\hl{{\cite{dahl2024large}}} and participants have reported a higher willingness to act on its advice. \hl{To this end a qualitative analysis of the advice provided, aided by legal experts, should be conducted.}

\hl{In this paper we have shown that participants report significantly higher willingness to act on legal advice generated by an LLM compared to when it is generated by a lawyer when the source of the advice is not known. However, while this finding in itself is interesting, it can not answer the `why is this the case'. While one might speculate that this could be linked to `undertrust in lawyers', `overtrust in AI', and `the specific use of language by either source' we can not know this for certain. Therefore, we suggest that future work extends on the here presented work by investigating, using qualitative data collection and analysis approaches, the underlying reasons for the here reported greater willingness to act on the LLM-generated advice.}

%% file: Sections/Conclusion.tex
\section{Conclusion}\label{sec:Conclusion}
Given the human-like nature of responses generated by LLMs, it becomes increasingly important to understand how lay people use this technology, especially in the context high-risk domains such as the legal context. In this paper, we have presented three experiments (total \textit{N}~=~288). Experiment 1 investigated if lay people are willing to act on legal advice---for advice on traffic, planning, and property law---when the source \hl{(lawyer- or LLM-generated)} of the advice was either known or unknown. Experiment 2 replicated the key manipulation of Experiment 1 using only the source unknown condition. Experiment 3 investigated if participants, when the source of advice was unknown, are able to discriminate the source. 

Findings of Experiment 1, successfully replicated in Experiment 2, show that participants, when the source of legal advice was unknown, report significantly higher willingness to act on the LLM-generated legal advice compared to the lawyer-generated advice. When the source of advice was known, no significant differences could be observed. Experiment 3 demonstrated, that even when participants were unaware of the source, they were able to discriminate the LLM- from the lawyer-generated advice significantly above chance \hl{(AUC = 0.59)}. Lastly, this paper discusses the importance of language used (e.g., complexity and advice length) when providing legal advice, possible causes for the change in ratings---going beyond the source---depending on if the source was known or not, risks associated with overtrust in LLMs and strategies to mitigate it, \hl{the need of LLMs to create---human and machine---identifiable content,} as well as limitations and future work.

%% file: Sections/Appendix.tex
\section{What is a Large Language Model (LLM)}\label{app:LLM}
The description presented to participants in all three experiments prior to the first case:
\begin{quote}
    \textit{`Large language models (LLMs) are computer programs that can read and generate human-like text by learning from vast amounts of written language. They can answer questions, write essays, or even create poetry, mimicking the style and content of the texts they were trained on. They can produce coherent and contextually relevant content, but their understanding is derived purely from the data they are trained on.'}~\cite{Seabrooke:2024:TAS}
\end{quote}